\documentclass[aps,jcp,preprint,showpacs,superscriptaddress,groupedaddress]{revtex4}  
\usepackage{graphicx}  
\usepackage{dcolumn}   
\usepackage{bm}        
\usepackage{amssymb}   
\usepackage{multirow}
\usepackage{tablefootnote}
\usepackage{times}
\usepackage{mathptm}
\usepackage[version=3]{mhchem}

\begin{document}

\title{The different facets of ice have different hydrophilicities: \\
  Friction at water / ice-I$_\mathrm{h}$ interfaces}
\author{Patrick B. Louden}
\author{J. Daniel Gezelter}
\email{gezelter@nd.edu}
\affiliation{Department of Chemistry and Biochemistry, University of
  Notre Dame, Notre Dame, IN 46556}

\date{\today}

\begin{abstract}
  We present evidence that the prismatic and secondary prism facets of
  ice-I$_\mathrm{h}$ crystals possess structural features that can
  reduce the effective hydrophilicity of the ice/water interface. The
  spreading dynamics of liquid water droplets on ice facets exhibits
  long-time behavior that differs for the prismatic
  $\{10\bar{1}0\}$ and secondary prism $\{11\bar{2}0\}$ facets
  when compared with the basal $\{0001\}$ and pyramidal
  $\{20\bar{2}1\}$ facets.  We also present the results of
  simulations of solid-liquid friction of the same four crystal facets
  being drawn through liquid water, and find that the two prismatic
  facets exhibit roughly half the solid-liquid friction of the basal
  and pyramidal facets.  These simulations provide evidence that the
  two prismatic faces have a significantly smaller effective surface
  area in contact with the liquid water. The ice / water interfacial
  widths for all four crystal facets are similar (using both
  structural and dynamic measures), and were found to be independent
  of the shear rate.  Additionally, decomposition of orientational
  time correlation functions show position-dependence for the short-
  and longer-time decay components close to the interface.
\end{abstract}

\pacs{68.08.Bc, 68.08.De, 66.20.Cy}
\keywords{ice; water; interfaces; hydrophobicity}
\maketitle

\section{Introduction}
Surfaces can be characterized as hydrophobic or hydrophilic
based on the strength of the interactions with water. Hydrophobic
surfaces do not have strong enough interactions with water to overcome
the internal attraction between molecules in the liquid phase, and the
degree of hydrophilicity of a surface can be described by the extent a
droplet can spread out over the surface. The contact angle, $\theta$,
formed between the solid and the liquid depends on the free energies
of the three interfaces involved, and is given by Young's
equation~\cite{Young05},
\begin{equation}\label{young}
\cos\theta = (\gamma_{sv} - \gamma_{sl})/\gamma_{lv} .
\end{equation} 
Here $\gamma_{sv}$, $\gamma_{sl}$, and $\gamma_{lv}$ are the free
energies of the solid/vapor, solid/liquid, and liquid/vapor interfaces,
respectively.  Large contact angles, $\theta > 90^{\circ}$, correspond
to hydrophobic surfaces with low wettability, while small contact
angles, $\theta < 90^{\circ}$, correspond to hydrophilic surfaces.
Experimentally, measurements of the contact angle of sessile drops is
often used to quantify the extent of wetting on surfaces with
thermally selective wetting
characteristics~\cite{Tadanaga00,Liu04,Sun04}.

Nanometer-scale structural features of a solid surface can influence
the hydrophilicity to a surprising degree.  Small changes in the
heights and widths of nano-pillars can change a surface from
superhydrophobic, $\theta \ge 150^{\circ}$, to hydrophilic, $\theta
\sim 0^{\circ}$~\cite{Koishi09}. This is often referred to as the
Cassie-Baxter to Wenzel transition.  Nano-pillared surfaces with
electrically tunable Cassie-Baxter and Wenzel states have also been
observed~\cite{Herbertson06,Dhindsa06,Verplanck07,Ahuja08,Manukyan11}.
Luzar and coworkers have modeled these transitions on nano-patterned
surfaces~\cite{Daub07,Daub10,Daub11,Ritchie12}, and have found the
change in contact angle is due to the field-induced perturbation of
hydrogen bonding at the liquid/vapor interface~\cite{Daub07}.

One would expect the interfaces of ice to be highly hydrophilic (and
possibly the most hydrophilic of all solid surfaces). In this paper we
present evidence that some of the crystal facets of ice-I$_\mathrm{h}$
have structural features that can reduce the effective hydrophilicity.
Our evidence for this comes from molecular dynamics (MD) simulations
of the spreading dynamics of liquid droplets on these facets, as well
as reverse non-equilibrium molecular dynamics (RNEMD) simulations of
solid-liquid friction.

Quiescent ice-I$_\mathrm{h}$/water interfaces have been studied
extensively using computer simulations. Hayward and Haymet
characterized and measured the widths of these
interfaces~\cite{Hayward01,Hayward02}.  Nada and Furukawa have also
modeled the width of basal/water and prismatic/water
interfaces~\cite{Nada95} as well as crystal restructuring at
temperatures approaching the melting point~\cite{Nada00}.

The surface of ice exhibits a pre-melting layer, often called a
quasi-liquid layer (QLL), at temperatures near the melting point.  MD
simulations of the facets of ice-I$_\mathrm{h}$ exposed to vacuum have
found QLL widths of approximately 10 \AA\ at 3 K below the melting
point~\cite{Conde08}. Similarly, Limmer and Chandler have used the mW
water model~\cite{Molinero09} and statistical field theory to estimate
QLL widths at similar temperatures to be about 3 nm~\cite{Limmer14}.

Recently, Sazaki and Furukawa have developed a technique using laser
confocal microscopy combined with differential interference contrast
microscopy that has sufficient spatial and temporal resolution to
visualize and quantitatively analyze QLLs on ice crystals at
temperatures near melting~\cite{Sazaki10}. They have found the width of
the QLLs perpendicular to the surface at -2.2$^{o}$C to be 3-4 \AA\
wide.  They have also seen the formation of two immiscible QLLs, which
displayed different dynamics on the crystal surface~\cite{Sazaki12}.


Using molecular dynamics simulations, Samadashvili has recently shown
that when two smooth ice slabs slide past one another, a stable
liquid-like layer develops between them~\cite{Samadashvili13}. In a
previous study, our RNEMD simulations of ice-I$_\mathrm{h}$ shearing
through liquid water have provided quantitative estimates of the
solid-liquid kinetic friction coefficients~\cite{Louden13}. These
displayed a factor of two difference between the basal and prismatic
facets.  The friction was found to be independent of shear direction
relative to the surface orientation.  We attributed facet-based
difference in liquid-solid friction to the 6.5 \AA\ corrugation of the
prismatic face which reduces the effective surface area of the ice
that is in direct contact with liquid water.

In the sections that follow, we describe the simulations of
droplet-spreading dynamics using standard MD as well as simulations of
tribological properties using RNEMD.  These simulations give
complementary results that point to the prismatic and secondary prism
facets having roughly half of their surface area in direct contact
with the liquid.

\section{Methodology}
\subsection{Construction of the Ice / Water interfaces}
Ice I$_\mathrm{h}$ crystallizes in the hexagonal space group
P$6_3/mmc$, and common ice crystals form hexagonal plates with the
basal face, $\{0001\}$, forming the top and bottom of each plate, and
the prismatic facet, $\{10\bar{1}0\}$, forming the sides.  In extreme
temperatures or low water saturation conditions, ice crystals can
easily form as hollow columns, needles and dendrites. These are
structures that expose other crystalline facets of the ice to the
surroundings.  Among the more common facets are the secondary prism,
$\{11\bar{2}0\}$, and pyramidal, $\{20\bar{2}1\}$, faces.  

We found it most useful to work with proton-ordered, zero-dipole
crystals that expose strips of dangling H-atoms and lone
pairs~\cite{Buch:2008fk}.  Our structures were created starting from
Structure 6 of Hirsch and Ojam\"{a}e's set of orthorhombic
representations for ice-I$_{h}$~\cite{Hirsch04}.  This crystal
structure was cleaved along the four different facets.  The exposed
face was reoriented normal to the $z$-axis of the simulation cell, and
the structures were extended to form large exposed facets in
rectangular box geometries.  Liquid water boxes were created with
identical dimensions (in $x$ and $y$) as the ice, with a $z$ dimension
of three times that of the ice block, and a density corresponding to 1
g / cm$^3$.  Each of the ice slabs and water boxes were independently
equilibrated at a pressure of 1 atm, and the resulting systems were
merged by carving out any liquid water molecules within 3 \AA\ of any
atoms in the ice slabs.  Each of the combined ice/water systems were
then equilibrated at 225K, which is the liquid-ice coexistence
temperature for SPC/E water~\cite{Bryk02}. Reference
\citealp{Louden13} contains a more detailed explanation of the
construction of similar ice/water interfaces. The resulting dimensions
as well as the number of ice and liquid water molecules contained in
each of these systems are shown in Table \ref{tab:method}.

The SPC/E water model~\cite{Berendsen87} has been extensively
characterized over a wide range of liquid
conditions~\cite{Arbuckle02,Kuang12}, and its phase diagram has been
well studied~\cite{Baez95,Bryk04b,Sanz04b,Fennell:2005fk}. With longer
cutoff radii and careful treatment of electrostatics, SPC/E mostly
avoids metastable crystalline morphologies like
ice-\textit{i}~\cite{Fennell:2005fk} and ice-B~\cite{Baez95}.  The
free energies and melting
points~\cite{Baez95,Arbuckle02,Gay02,Bryk02,Bryk04b,Sanz04b,Fennell:2005fk,Fernandez06,Abascal07,Vrbka07}
of various other crystalline polymorphs have also been calculated.
Haymet \textit{et al.} have studied quiescent Ice-I$_\mathrm{h}$/water
interfaces using the SPC/E water model, and have seen structural and
dynamic measurements of the interfacial width that agree well with
more expensive water models, although the coexistence temperature for
SPC/E is still well below the experimental melting point of real
water~\cite{Bryk02}. Given the extensive data and speed of this model,
it is a reasonable choice even though the temperatures required are
somewhat lower than real ice / water interfaces.

\section{Droplet Simulations}
\begin{figure}
\includegraphics[width=\linewidth]{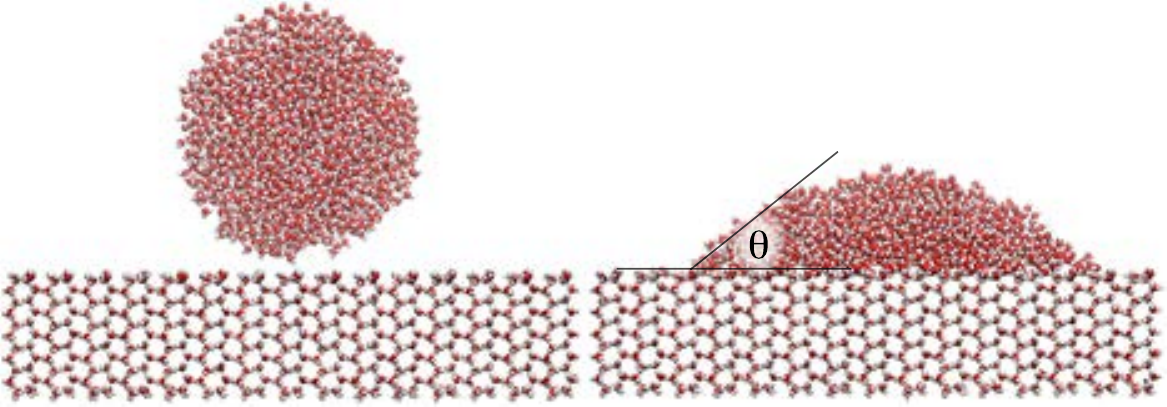}
\caption{\label{fig:Droplet} Computational model of a droplet of
  liquid water spreading over the prismatic $\{10\bar{1}0\}$ facet
  of ice, before (left) and 2.6 ns after (right) being introduced to the
  surface.  The contact angle ($\theta$) shrinks as the simulation
  proceeds, and the long-time behavior of this angle is used to
  estimate the hydrophilicity of the facet.}
\end{figure}

Ice surfaces with a thickness of $\sim$~20~\AA\ were created as
described above, but were not solvated in a liquid box. The crystals
were then replicated along the $x$ and $y$ axes (parallel to the
surface) until a large surface ($>$ 126 nm\textsuperscript{2}) had
been created.  The sizes and numbers of molecules in each of the
surfaces is given in Table S1.  Weak translational restraining
potentials with spring constants of 1.5~$\mathrm{kcal\
  mol}^{-1}\mathrm{~\AA}^{-2}$ (prismatic and pyramidal facets) or
4.0~$\mathrm{kcal\ mol}^{-1}\mathrm{~\AA}^{-2}$ (basal facet) were
applied to the centers of mass of each molecule in order to prevent
surface melting, although the molecules were allowed to reorient
freely. A water droplet containing 2048 SPC/E molecules was created
separately. Droplets of this size can produce agreement with the Young
contact angle extrapolated to an infinite drop size~\cite{Daub10}. The
surfaces and droplet were independently equilibrated to 225~K, at
which time the droplet was placed 3-5~\AA\ above the surface.  Five
statistically independent simulations were carried out for each facet,
and the droplet was placed at unique $x$ and $y$ locations for each of
these simulations.  Each simulation was 5~ns in length and was
conducted in the microcanonical (NVE) ensemble.  Representative
configurations for the droplet on the prismatic facet are shown in
figure \ref{fig:Droplet}.

\section{Shearing Simulations (Interfaces in Bulk Water)}
To perform the shearing simulations, the velocity shearing and scaling
variant of reverse non-equilibrium molecular dynamics (VSS-RNEMD) was
employed \cite{Kuang12}. This method performs a series of simultaneous
non-equilibrium exchanges of linear momentum and kinetic energy
between two physically-separated regions of the simulation cell.  The
system responds to this unphysical flux with velocity and temperature
gradients.  When VSS-RNEMD is applied to bulk liquids, transport
properties like the thermal conductivity and the shear viscosity are
easily extracted assuming a linear response between the flux and the
gradient.  At the interfaces between dissimilar materials, the same
method can be used to extract \textit{interfacial} transport
properties (e.g. the interfacial thermal conductance and the
hydrodynamic slip length).

\begin{figure}
\includegraphics[width=1.9in]{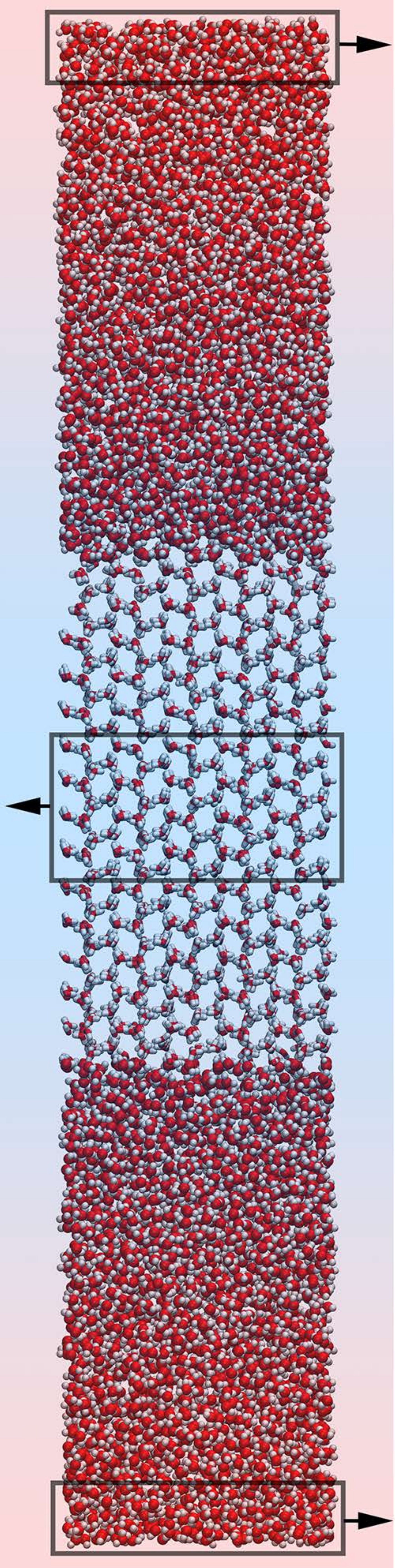}
\caption{\label{fig:Shearing} Computational model of a slab of ice
  being sheared through liquid water.  In this figure, the ice is
  presenting two copies of the prismatic $\{10\bar{1}0\}$ facet
  towards the liquid phase.  The RNEMD simulation exchanges both
  linear momentum (indicated with arrows) and kinetic energy between
  the central box and the box that spans the cell boundary.  The
  system responds with weak thermal gradient and a velocity profile
  that shears the ice relative to the surrounding liquid.}
\end{figure}

The kinetic energy flux (producing a thermal gradient) is necessary
when performing shearing simulations at the ice-water interface in
order to prevent the frictional heating due to the shear from melting
the crystal. Reference \citealp{Louden13} provides more details on the
VSS-RNEMD method as applied to ice-water interfaces. A representative
configuration of the solvated prismatic facet being sheared through
liquid water is shown in figure \ref{fig:Shearing}.

All simulations were performed using OpenMD~\cite{OOPSE,openmd}, with
a time step of 2 fs and periodic boundary conditions in all three
dimensions.  Electrostatics were handled using the damped-shifted
force real-space electrostatic kernel~\cite{Ewald}. 

The interfaces were equilibrated for a total of 10 ns at equilibrium
conditions before being exposed to either a shear or thermal gradient.
This consisted of 5 ns under a constant temperature (NVT) integrator
set to 225~K followed by 5 ns under a microcanonical (NVE) integrator.
Weak thermal gradients were allowed to develop using the VSS-RNEMD
(NVE) integrator using a small thermal flux ($-2.0\times 10^{-6}$
kcal/mol/\AA$^2$/fs) for a duration of 5 ns to allow the gradient to
stabilize.  The resulting temperature gradient was $\approx$ 10K over
the entire box length, which was sufficient to keep the temperature at
the interface within $\pm 1$ K of the 225~K target.

Velocity gradients were then imposed using the VSS-RNEMD (NVE)
integrator with a range of momentum fluxes.  The systems were divided
into 100 bins along the $z$-axis for the VSS-RNEMD moves, which were
attempted every time step.  Although computationally expensive, this
was done to minimize the magnitude of each individual momentum
exchange.  Because individual VSS-RNEMD exchange moves conserve both
total energy and linear momentum, the method can be ``bolted-on'' to
simulations in any ensemble.  The simulations of the pyramidal
interface were performed under the canonical (NVT) ensemble.  When
time correlation functions were computed, the RNEMD simulations were
done in the microcanonical (NVE) ensemble.  All simulations of the
other interfaces were carried out in the microcanonical ensemble.

These gradients were allowed to stabilize for 1~ns before data
collection started. Once established, four successive 0.5~ns runs were
performed for each shear rate.  During these simulations,
configurations of the system were stored every 1~ps, and statistics on
the structure and dynamics in each bin were accumulated throughout the
simulations.  Although there was some small variation in the measured
interfacial width between succcessive runs, no indication of bulk
melting or crystallization was observed.  That is, no large scale
changes in the positions of the top and bottom interfaces occurred
during the simulations.

\section{Results}
\subsection{Ice - Water Contact Angles}
To determine the extent of wetting for each of the four crystal
facets, contact angles for liquid droplets on the ice surfaces were
computed using two methods.  In the first method, the droplet is
assumed to form a spherical cap, and the contact angle is estimated
from the $z$-axis location of the droplet's center of mass
($z_\mathrm{cm}$).  This procedure was first described by Hautman and
Klein~\cite{Hautman91}, and was utilized by Hirvi and Pakkanen in
their investigation of water droplets on polyethylene and poly(vinyl
chloride) surfaces~\cite{Hirvi06}. For each stored configuration, the
contact angle, $\theta$, was found by inverting the expression for the
location of the droplet center of mass,
\begin{equation}\label{contact_1}
\langle z_\mathrm{cm}\rangle = 2^{-4/3}R_{0}\bigg(\frac{1-cos\theta}{2+cos\theta}\bigg)^{1/3}\frac{3+cos\theta}{2+cos\theta} ,
\end{equation}
where $R_{0}$ is the radius of the free water droplet. 

In addition to the spherical cap method outlined above, a second
method for obtaining the contact angle was described by Ruijter,
Blake, and Coninck~\cite{Ruijter99}.  This method uses a cylindrical
averaging of the droplet's density profile.  A threshold density of
0.5 g cm\textsuperscript{-3} is used to estimate the location of the
edge of the droplet.  The $r$ and $z$-dependence of the droplet's edge
is then fit to a circle, and the contact angle is computed from the
intersection of the fit circle with the $z$-axis location of the solid
surface.  Again, for each stored configuration, the density profile in
a set of annular shells was computed. Due to large density
fluctuations close to the ice, all shells located within 2 \AA\ of the
ice surface were left out of the circular fits.  The height of the
solid surface ($z_\mathrm{suface}$) along with the best fitting origin
($z_\mathrm{droplet}$) and radius ($r_\mathrm{droplet}$) of the
droplet can then be used to compute the contact angle,
\begin{equation}
\theta =  90 + \frac{180}{\pi} \sin^{-1}\left(\frac{z_\mathrm{droplet} -
  z_\mathrm{surface}}{r_\mathrm{droplet}} \right).
\end{equation}
Both methods provided similar estimates of the dynamic contact angle,
although the spherical cap method is significantly less prone to
noise, and is the method used to compute the contact angles in table
\ref{tab:kappa}.

Because the initial droplet was placed above the surface, the initial
value of 180$^{\circ}$ decayed over time (See figure 1 in the
SI).  Each of these profiles were fit to a
biexponential decay, with a short-time contribution ($\tau_c$) that
describes the initial contact with the surface, a long time
contribution ($\tau_s$) that describes the spread of the droplet over
the surface, and a constant ($\theta_\infty$) to capture the
infinite-time estimate of the equilibrium contact angle,
\begin{equation}
\theta(t) = \theta_\infty +  (180-\theta_\infty) \left[ a e^{-t/\tau_c} +
  (1-a) e^{-t/\tau_s}  \right]
\end{equation}
We have found that the rate for water droplet spreading across all
four crystal facets, $k_\mathrm{spread} = 1/\tau_s \approx$ 0.7
ns$^{-1}$. However, the basal and pyramidal facets produced estimated
equilibrium contact angles, $\theta_\infty \approx$ 35$^{\circ}$, while
prismatic and secondary prismatic had values for $\theta_\infty$ near
43$^{\circ}$ as seen in Table \ref{tab:kappa}.

These results indicate that by traditional measures, the basal and
pyramidal facets are more hydrophilic than the prismatic and secondary
prism facets, and surprisingly, that the differential hydrophilicities
of the crystal facets is not reflected in the spreading rate of the
droplet.


\subsection{Solid-liquid friction of the interfaces}
In a bulk fluid, the shear viscosity, $\eta$, can be determined
assuming a linear response to a shear stress,
\begin{equation}\label{Shenyu-11}
j_{z}(p_{x}) = \eta \frac{\partial v_{x}}{\partial z}.
\end{equation}
Here $j_{z}(p_{x})$ is the flux (in $x$-momentum) that is transferred
in the $z$ direction (i.e. the shear stress). The RNEMD simulations
impose an artificial momentum flux between two regions of the
simulation, and the velocity gradient is the fluid's response. This
technique has now been applied quite widely to determine the
viscosities of a number of bulk fluids~\cite{Muller99,Bordat02,Cavalcanti07}.

At the interface between two phases (e.g. liquid / solid) the same
momentum flux creates a velocity difference between the two materials,
and this can be used to define an interfacial friction coefficient
($\kappa$),
\begin{equation}\label{Shenyu-13}
j_{z}(p_{x}) = \kappa \left[ v_{x}(liquid) - v_{x}(solid) \right]
\end{equation}
where $v_{x}(solid)$ is the velocity of the solid and $v_{x}(liquid)$
is the velocity of the liquid measured at the hydrodynamic boundary
layer.

The simulations described here contain significant quantities of both
liquid and solid phases, and the momentum flux must traverse a region
of the liquid that is simultaneously under a thermal gradient.  Since
the liquid has a temperature-dependent shear viscosity, $\eta(T)$,
estimates of the solid-liquid friction coefficient can be obtained if
one knows the viscosity of the liquid at the interface (i.e. at the
melting temperature, $T_m$),
\begin{equation}\label{kappa-2}
\kappa = \frac{\eta(T_{m})}{\left[v_{x}(fluid)-v_{x}(solid)\right]}\left(\frac{\partial v_{x}}{\partial z}\right).
\end{equation}
For SPC/E, the melting temperature of Ice-I$_\mathrm{h}$ is estimated
to be 225~K~\cite{Bryk02}.  To obtain the value of
$\eta(225\mathrm{~K})$ for the SPC/E model, a $31.09 \times 29.38
\times 124.39$ \AA\ box with 3744 water molecules in a disordered
configuration was equilibrated to 225~K, and five
statistically-independent shearing simulations were performed (with
imposed fluxes that spanned a range of $3 \rightarrow 13
\mathrm{~m~s}^{-1}$ ).  Each simulation was conducted in the
microcanonical ensemble with total simulation times of 5 ns. The
VSS-RNEMD exchanges were carried out every 2 fs. We estimate
$\eta(225\mathrm{~K})$ to be 0.0148 $\pm$ 0.0007 Pa s for SPC/E,
roughly ten times larger than the shear viscosity previously computed
at 280~K~\cite{Kuang12}.

The interfacial friction coefficient can equivalently be expressed as
the ratio of the viscosity of the fluid to the hydrodynamic slip
length, $\kappa = \eta / \delta$. The slip length is an indication of
strength of the interactions between the solid and liquid phases,
although the connection between slip length and surface hydrophobicity
is not yet clear. In some simulations, the slip length has been found
to have a link to the effective surface
hydrophobicity~\cite{Sendner:2009uq}, although Ho \textit{et al.} have
found that liquid water can also slip on hydrophilic
surfaces~\cite{Ho:2011zr}. Experimental evidence for a direct tie
between slip length and hydrophobicity is also not
definitive. Total-internal reflection particle image velocimetry
(TIR-PIV) studies have suggested that there is a link between slip
length and effective
hydrophobicity~\cite{Lasne:2008vn,Bouzigues:2008ys}. However, recent
surface sensitive cross-correlation spectroscopy (TIR-FCCS)
measurements have seen similar slip behavior for both hydrophobic and
hydrophilic surfaces~\cite{Schaeffel:2013kx}.

In each of the systems studied here, the interfacial temperature was
kept fixed to 225~K, which ensured the viscosity of the fluid at the
interace was identical. Thus, any significant variation in $\kappa$
between the systems is a direct indicator of the slip length and the
effective interaction strength between the solid and liquid phases.

The calculated $\kappa$ values found for the four crystal facets of
Ice-I$_\mathrm{h}$ are shown in Table \ref{tab:kappa}. The basal and
pyramidal facets were found to have similar values of $\kappa \approx
6$ ($\times 10^{-4} \mathrm{amu~\AA}^{-2}\mathrm{fs}^{-1}$), while the
prismatic and secondary prism facets exhibited $\kappa \approx 3$
($\times 10^{-4} \mathrm{amu~\AA}^{-2}\mathrm{fs}^{-1}$). These
results are also essentially independent of the direction of the shear
relative to channels on the surfaces of the facets.  The friction
coefficients indicate that the basal and pyramidal facets have
significantly stronger interactions with liquid water than either of
the two prismatic facets.  This is in agreement with the contact angle
results above - both of the high-friction facets exhibited smaller
contact angles, suggesting that the solid-liquid friction (and inverse
slip length) is correlated with the hydrophilicity of these facets.

\subsection{Structural measures of interfacial width under shear}
One of the open questions about ice/water interfaces is whether the
thickness of the 'slush-like' quasi-liquid layer (QLL) depends on the
facet of ice presented to the water.  In the QLL region, the water
molecules are ordered differently than in either the solid or liquid
phases, and also exhibit distinct dynamical behavior.  The width of
this quasi-liquid layer has been estimated by finding the distance
over which structural order parameters or dynamic properties change
from their bulk liquid values to those of the solid ice.  The
properties used to find interfacial widths have included the local
density, the diffusion constant, and both translational and
orientational order
parameters~\cite{Karim88,Karim90,Hayward01,Hayward02,Bryk02,Gay02,Louden13}.

The VSS-RNEMD simulations impose thermal and velocity gradients.
These gradients perturb the momenta of the water molecules, so
parameters that depend on translational motion are often measuring the
momentum exchange, and not physical properties of the interface.  As a
structural measure of the interface, we have used the local
tetrahedral order parameter, which measures the match of the local
molecular environments (e.g. the angles between nearest neighbor
molecules) to perfect tetrahedral ordering.  This quantity was
originally described by Errington and Debenedetti~\cite{Errington01}
and has been used in bulk simulations by Kumar \textit{et
  al.}~\cite{Kumar09} It has previously been used in ice/water
interfaces by by Bryk and Haymet~\cite{Bryk04b}.

To determine the structural widths of the interfaces under shear, each
of the systems was divided into 100 bins along the $z$-dimension, and
the local tetrahedral order parameter (Eq. 5 in Reference
\citealp{Louden13}) was time-averaged in each bin for the duration of
the shearing simulation.  The spatial dependence of this order
parameter, $q(z)$, is the tetrahedrality profile of the interface.
The lower panels in figures 2-5 in the supporting information show
tetrahedrality profiles (in circles) for each of the four interfaces.
The $q(z)$ function has a range of $(0,1)$, where a value of unity
indicates a perfectly tetrahedral environment.  The $q(z)$ for the
bulk liquid was found to be $\approx~0.77$, while values of
$\approx~0.92$ were more common in the ice. The tetrahedrality
profiles were fit using a hyperbolic tangent function (see Eq. 6 in
Reference \citealp{Louden13}) designed to smoothly fit the bulk to ice
transition while accounting for the weak thermal gradient. In panels
$b$ and $c$ of the same figures, the resulting thermal and velocity
gradients from an imposed kinetic energy and momentum fluxes can be
seen. The vertical dotted lines traversing these figures indicate the
midpoints of the interfaces as determined by the tetrahedrality
profiles.  The hyperbolic tangent fit provides an estimate of
$d_\mathrm{struct}$, the structural width of the interface.
 
We find the interfacial width to be $3.2 \pm 0.2$ \AA\ (pyramidal) and
$3.2 \pm 0.2$ \AA\ (secondary prism) for the control systems with no
applied momentum flux. This is similar to our previous results for the
interfacial widths of the quiescent basal ($3.2 \pm 0.4$ \AA) and
prismatic systems ($3.6 \pm 0.2$ \AA).

Over the range of shear rates investigated, $0.4 \rightarrow
6.0~\mathrm{~m~s}^{-1}$ for the pyramidal system and $0.6 \rightarrow
5.2~\mathrm{~m~s}^{-1}$ for the secondary prism, we found no
significant change in the interfacial width. The mean interfacial
widths are collected in table \ref{tab:kappa}. This follows our
previous findings of the basal and prismatic systems, in which the
interfacial widths of the basal and prismatic facets were also found
to be insensitive to the shear rate~\cite{Louden13}.

The similarity of these interfacial width estimates indicate that the
particular facet of the exposed ice crystal has little to no effect on
how far into the bulk the ice-like structural ordering persists. Also,
it appears that for the shearing rates imposed in this study, the
interfacial width is not structurally modified by the movement of
water over the ice.

\subsection{Dynamic measures of interfacial width under shear}
The spatially-resolved orientational time correlation function,
\begin{equation}\label{C(t)1}
  C_{2}(z,t)=\langle P_{2}(\mathbf{u}_i(0)\cdot \mathbf{u}_i(t))
  \delta(z_i(0) - z) \rangle,
\end{equation}
provides local information about the decorrelation of molecular
orientations in time. Here, $P_{2}$ is the second-order Legendre
polynomial, and $\mathbf{u}_i$ is the molecular vector that bisects
the HOH angle of molecule $i$.  The angle brackets indicate an average
over all the water molecules and time origins, and the delta function
restricts the average to specific regions. In the crystal, decay of
$C_2(z,t)$ is incomplete, while in the liquid, correlation times are
typically measured in ps.  Observing the spatial-transition between
the decay regimes can define a dynamic measure of the interfacial
width.

To determine the dynamic widths of the interfaces under shear, each of
the systems was divided into bins along the $z$-dimension ($\approx$ 3
\AA\ wide) and $C_2(z,t)$ was computed using only those molecules that
were in the bin at the initial time.  To compute these correlation
functions, each of the 0.5 ns simulations was followed by a shorter
200 ps microcanonical (NVE) simulation in which the positions and
orientations of every molecule in the system were recorded every 0.1
ps. 

The time-dependence was fit to a triexponential decay, with three time
constants: $\tau_{short}$, measuring the librational motion of the
water molecules, $\tau_{middle}$, measuring the timescale for breaking
and making of hydrogen bonds, and $\tau_{long}$, corresponding to the
translational motion of the water molecules.  An additional constant
was introduced in the fits to describe molecules in the crystal which
do not experience long-time orientational decay.

In Figures 6-9 in the supporting information, the $z$-coordinate
profiles for the three decay constants, $\tau_{short}$,
$\tau_{middle}$, and $\tau_{long}$ for the different interfaces are
shown.  (Figures 6 \& 7 are new results, and Figures 8 \& 9 are
updated plots from Ref \citealp{Louden13}.)  In the liquid regions of
all four interfaces, we observe $\tau_{middle}$ and $\tau_{long}$ to
have approximately consistent values of $3-6$ ps and $30-40$ ps,
respectively.  Both of these times increase in value approaching the
interface.  Approaching the interface, we also observe that
$\tau_{short}$ decreases from its liquid-state value of $72-76$ fs.
The approximate values for the decay constants and the trends
approaching the interface match those reported previously for the
basal and prismatic interfaces.

We have estimated the dynamic interfacial width $d_\mathrm{dyn}$ by
fitting the profiles of all the three orientational time constants
with an exponential decay to the bulk-liquid behavior,
\begin{equation}\label{tauFit}
  \tau(z)\approx\tau_{liquid}+(\tau_{wall}-\tau_{liquid})e^{-(z-z_{wall})/d_\mathrm{dyn}}
\end{equation}
where $\tau_{liquid}$ and $\tau_{wall}$ are the liquid and projected
wall values of the decay constants, $z_{wall}$ is the location of the
interface, as measured by the structural order parameter.  These
values are shown in table \ref{tab:kappa}. Because the bins must be
quite wide to obtain reasonable profiles of $C_2(z,t)$, the error
estimates for the dynamic widths of the interface are significantly
larger than for the structural widths.  However, all four interfaces
exhibit dynamic widths that are significantly below 1~nm, and are in
reasonable agreement with the structural width above.

\section{Conclusions}
In this work, we used MD simulations to measure the advancing contact
angles of water droplets on the basal, prismatic, pyramidal, and
secondary prism facets of Ice-I$_\mathrm{h}$.  Although we saw no
significant change in the \textit{rate} at which the droplets spread
over the surface, the long-time behavior predicts larger equilibrium
contact angles for the two prismatic facets.  

We have also used RNEMD simulations of water interfaces with the same
four crystal facets to compute solid-liquid friction coefficients.  We
have observed coefficients of friction that differ by a factor of two
between the two prismatic facets and the basal and pyramidal facets.
Because the solid-liquid friction coefficient is directly tied to the
inverse of the hydrodynamic slip length, this suggests that there are
significant differences in the overall interaction strengths between
these facets and the liquid layers immediately in contact with them.

The agreement between these two measures have lead us to conclude that
the two prismatic facets have a lower hydrophilicity than either the
basal or pyramidal facets.  One possible explanation of this behavior
is that the face presented by both prismatic facets consists of deep,
narrow channels (i.e. stripes of adjacent rows of pairs of
hydrogen-bound water molecules).  At the surfaces of these facets,
the channels are 6.35 \AA\ wide and the sub-surface ice layer is 2.25
\AA\ below (and therefore blocked from hydrogen bonding with the
liquid).  This means that only 1/2 of the surface molecules can form
hydrogen bonds with liquid-phase molecules.

In the basal plane, the surface features are shallower (1.3 \AA), with
no blocked subsurface layer.  The pyramidal face has much wider
channels (8.65 \AA) which are also quite shallow (1.37 \AA).  These
features allow liquid phase molecules to form hydrogen bonds with all
of the surface molecules in the basal and pyramidal facets.  This
means that for similar surface areas, the two prismatic facets have an
effective hydrogen bonding surface area of half of the basal and
pyramidal facets.  The reduction in the effective surface area would
explain much of the behavior observed in our simulations.

\begin{acknowledgments}
  Support for this project was provided by the National
  Science Foundation under grant CHE-1362211. Computational time was
  provided by the Center for Research Computing (CRC) at the
  University of Notre Dame.
\end{acknowledgments}
\bibliographystyle{aip}
\newpage
\bibliography{iceWater}
\newpage
\newpage

\newpage
\begin{table}[h]
\centering
\caption{Sizes of the droplet and shearing simulations.  Cell
  dimensions are measured in \AA. \label{tab:method}}
\begin{tabular}{r|cccc|ccccc} 
\toprule
 \multirow{2}{*}{Interface} & \multicolumn{4}{c|}{Droplet} & \multicolumn{5}{c}{
Shearing} \\
  & $N_\mathrm{ice}$ & $N_\mathrm{droplet}$ & $L_x$ & $L_y$ & $N_\mathrm{ice}$ &
 $N_\mathrm{liquid}$ & $L_x$ & $L_y$ & $L_z$  \\ 
\colrule
Basal  $\{0001\}$                    & 12960 & 2048 & 134.70 & 140.04 & 900 & 1846  & 23.87 & 35.83 & 98.64  \\
Pyramidal  $\{20\bar{2}1\}$       & 11136 & 2048 & 143.75 & 121.41 & 1216 & 2203 & 37.47 & 29.50 & 93.02  \\
Prismatic  $\{10\bar{1}0\}$       &  9900 & 2048 & 110.04 & 115.00 & 3000 & 5464 & 35.95 & 35.65 & 205.77 \\
Secondary Prism  $\{11\bar{2}0\}$ & 11520 & 2048 & 146.72 & 124.48 & 3840 & 8176 & 71.87 & 31.66 & 161.55 \\
\botrule
\end{tabular}
\end{table}

\newpage
\begin{table}[h]
\centering
\caption{Structural and dynamic properties of the interfaces of
  Ice-I$_\mathrm{h}$ with water.\label{tab:kappa}}
\begin{tabular}{r|cc|cc|cccc}  
\toprule
\multirow{2}{*}{Interface} & \multicolumn{2}{c|}{Channel Size} &\multicolumn{2}{c|}{Droplet} & \multicolumn{4}{c}{Shearing\footnotemark[1]}\\
  & Width (\AA) & Depth (\AA) & $\theta_{\infty}$ ($^\circ$)  & $k_\mathrm{spread}$  (ns\textsuperscript{-1}) &
$\kappa_{x}$  & $\kappa_{y}$ & $d_\mathrm{struct}$ (\AA) &  $d_\mathrm{dyn}$ (\AA) \\ 
\colrule
Basal  $\{0001\}$                    & 4.49 & 1.30 & $34.1(9)$ &$0.60(7)$
& $5.9(3)$ & $6.5(8)$ & $3.2(4)$ & $2(1)$  \\
Pyramidal  $\{20\bar{2}1\}$       & 8.65 & 1.37 & $35(3)$ &  $0.7(1)$ &
$5.8(4)$ & $6.1(5)$ & $3.2(2)$ & $2.5(3)$\\
Prismatic  $\{10\bar{1}0\}$       & 6.35 & 2.25 & $45(3)$ & $0.75(9)$ &
$3.0(2)$ & $3.0(1)$ & $3.6(2)$ & $4(2)$ \\
Secondary Prism  $\{11\bar{2}0\}$ & 6.35 & 2.25 & $43(2)$ & $0.69(3)$ &
$3.5(1)$ & $3.3(2)$ & $3.2(2)$ & $5(3)$ \\ 
\botrule
\end{tabular}
\begin{flushleft}
\footnotemark[1]\footnotesize{Liquid-solid friction coefficients ($\kappa_x$ and
  $\kappa_y$) are expressed in 10\textsuperscript{-4} amu
  \AA\textsuperscript{-2} fs\textsuperscript{-1}.} \\
\footnotemark[2]\footnotesize{Uncertainties in
  the last digits are given in parentheses.} 
\end{flushleft}
\end{table}


\end{document}


\title{Supporting Information for: \\
The different facets of ice have different hydrophilicities: Friction at water /
  ice-I$_\mathrm{h}$ interfaces}

\author{Patrick B. Louden}
\author{J. Daniel Gezelter}
\email{gezelter@nd.edu}
\affiliation{Department of Chemistry and Biochemistry, University of
  Notre Dame, Notre Dame, IN 46556}

\date{\today}

\begin{abstract}
The supporting information supplies figures that support the data
presented in the main text.
\end{abstract}

\pacs{68.08.Bc, 68.08.De, 66.20.Cy}

\maketitle

\section{The Advancing Contact Angle}
The advancing contact angles for the liquid droplets were computed
using inversion of Eq. (2) in the main text which requires finding the
real roots of a fourth order polynomial,
\begin{equation}
\label{eq:poly}
c_4 \cos^4 \theta + c_3 \cos^3 \theta + c_2 \cos^2 \theta + c_1
\cos \theta + c_0 = 0
\end{equation}
where the coefficients of the polynomial are expressed in terms of the
$z$ coordinate of the center of mass of the liquid droplet relative to
the solid surface, $z = z_\mathrm{cm} - z_\mathrm{surface}$, and a
factor that depends on the initial droplet radius, $k = 2^{-4/3} R_0$.
The coefficients are simple functions of these two quantities,
\begin{align}
c_4 &= z^3 + k^3 \\
c_3 &= 8 z^3 + 8 k^3 \\
c_2 &= 24 z^3 + 18 k^3 \\
c_1 &= 32 z^3 \\
c_0 &= 16 z^3 - 27 k^3 .
\end{align}
Solving for the values of the real roots of this polynomial
(Eq. \ref{eq:poly}) give estimates of the advancing contact angle.
The dynamics of this quantity for each of the four interfaces is shown
in figure 1 below.

\section{Interfacial widths using structural information}
To determine the structural widths of the interfaces under shear, each
of the systems was divided into 100 bins along the $z$-dimension, and
the local tetrahedral order parameter (Eq. 5 in Reference
\citealp{Louden13}) was time-averaged in each bin for the duration of
the shearing simulation.  The spatial dependence of this order
parameter, $q(z)$, is the tetrahedrality profile of the interface.
The lower panels in figures 2-5 show tetrahedrality profiles (in
circles) for each of the four interfaces.  The $q(z)$ function has a
range of $(0,1)$, where a value of unity indicates a perfectly
tetrahedral environment.  The $q(z)$ for the bulk liquid was found to
be $\approx~0.77$, while values of $\approx~0.92$ were more common in
the ice. The tetrahedrality profiles were fit using a hyperbolic
tangent function (see Eq. 6 in Reference \citealp{Louden13}) designed
to smoothly fit the bulk to ice transition while accounting for the
weak thermal gradient. In panels $b$ and $c$ of the same figures, the
resulting thermal and velocity gradients from an imposed kinetic
energy and momentum fluxes can be seen. The vertical dotted lines
traversing these figures indicate the midpoints of the interfaces as
determined by the tetrahedrality profiles.

\section{Interfacial widths using dynamic information}
To determine the dynamic widths of the interfaces under shear, each of
the systems was divided into bins along the $z$-dimension ($\approx$ 3
\AA\ wide) and $C_2(z,t)$ was computed using only those molecules that
were in the bin at the initial time.  To compute these correlation
functions, each of the 0.5 ns simulations was followed by a shorter
200 ps microcanonical (NVE) simulation in which the positions and
orientations of every molecule in the system were recorded every 0.1
ps. 

The time-dependence was fit to a triexponential decay, with three time
constants: $\tau_{short}$, measuring the librational motion of the
water molecules, $\tau_{middle}$, measuring the timescale for breaking
and making of hydrogen bonds, and $\tau_{long}$, corresponding to the
translational motion of the water molecules.  An additional constant
was introduced in the fits to describe molecules in the crystal which
do not experience long-time orientational decay.

In Figures 6-9, the $z$-coordinate profiles for the three decay
constants, $\tau_{short}$, $\tau_{middle}$, and $\tau_{long}$ for the
different interfaces are shown.  (Figures 6 \& 7 are new results,
and Figures 8 \& 9 are updated plots from Ref \citealp{Louden13}.)
In the liquid regions of all four interfaces, we observe
$\tau_{middle}$ and $\tau_{long}$ to have approximately consistent
values of $3-6$ ps and $30-40$ ps, respectively.  Both of these times
increase in value approaching the interface.  Approaching the
interface, we also observe that $\tau_{short}$ decreases from its
liquid-state value of $72-76$ fs.  The approximate values for the
decay constants and the trends approaching the interface match those
reported previously for the basal and prismatic interfaces.

We have estimated the dynamic interfacial width $d_\mathrm{dyn}$ by
fitting the profiles of all the three orientational time constants
with an exponential decay to the bulk-liquid behavior,
\begin{equation}\label{tauFit}
  \tau(z)\approx\tau_{liquid}+(\tau_{wall}-\tau_{liquid})e^{-(z-z_{wall})/d_\mathrm{dyn}}
\end{equation}
where $\tau_{liquid}$ and $\tau_{wall}$ are the liquid and projected
wall values of the decay constants, $z_{wall}$ is the location of the
interface, as measured by the structural order parameter.  These
values are shown in table 1 in the main text. Because the bins must be
quite wide to obtain reasonable profiles of $C_2(z,t)$, the error
estimates for the dynamic widths of the interface are significantly
larger than for the structural widths.  However, all four interfaces
exhibit dynamic widths that are significantly below 1~nm, and are in
reasonable agreement with the structural width above.

\bibliographystyle{aip}
\bibliography{iceWater}

\newpage
\begin{figure}
\includegraphics[width=\linewidth]{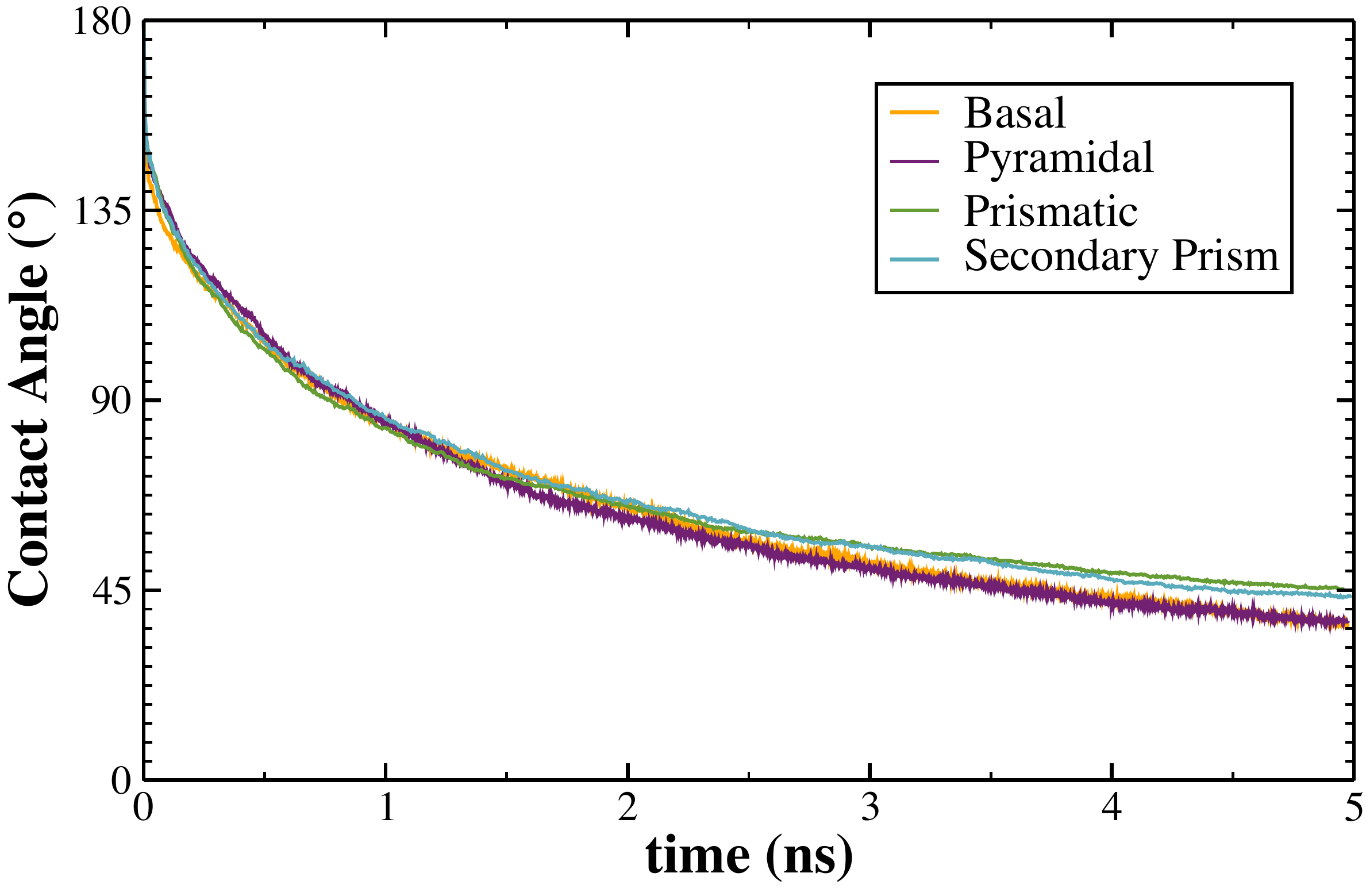}
\caption{\label{fig:ContactAngle} The dynamic contact angle of a
  droplet after approaching each of the four ice facets.  The decay to
  an equilibrium contact angle displays similar dynamics.  Although
  all the surfaces are hydrophilic, the long-time behavior stabilizes
  to significantly flatter droplets for the basal and pyramidal
  facets.  This suggests a difference in hydrophilicity for these
  facets compared with the two prismatic facets.}
\end{figure}

\newpage

\begin{figure}
\includegraphics[width=\linewidth]{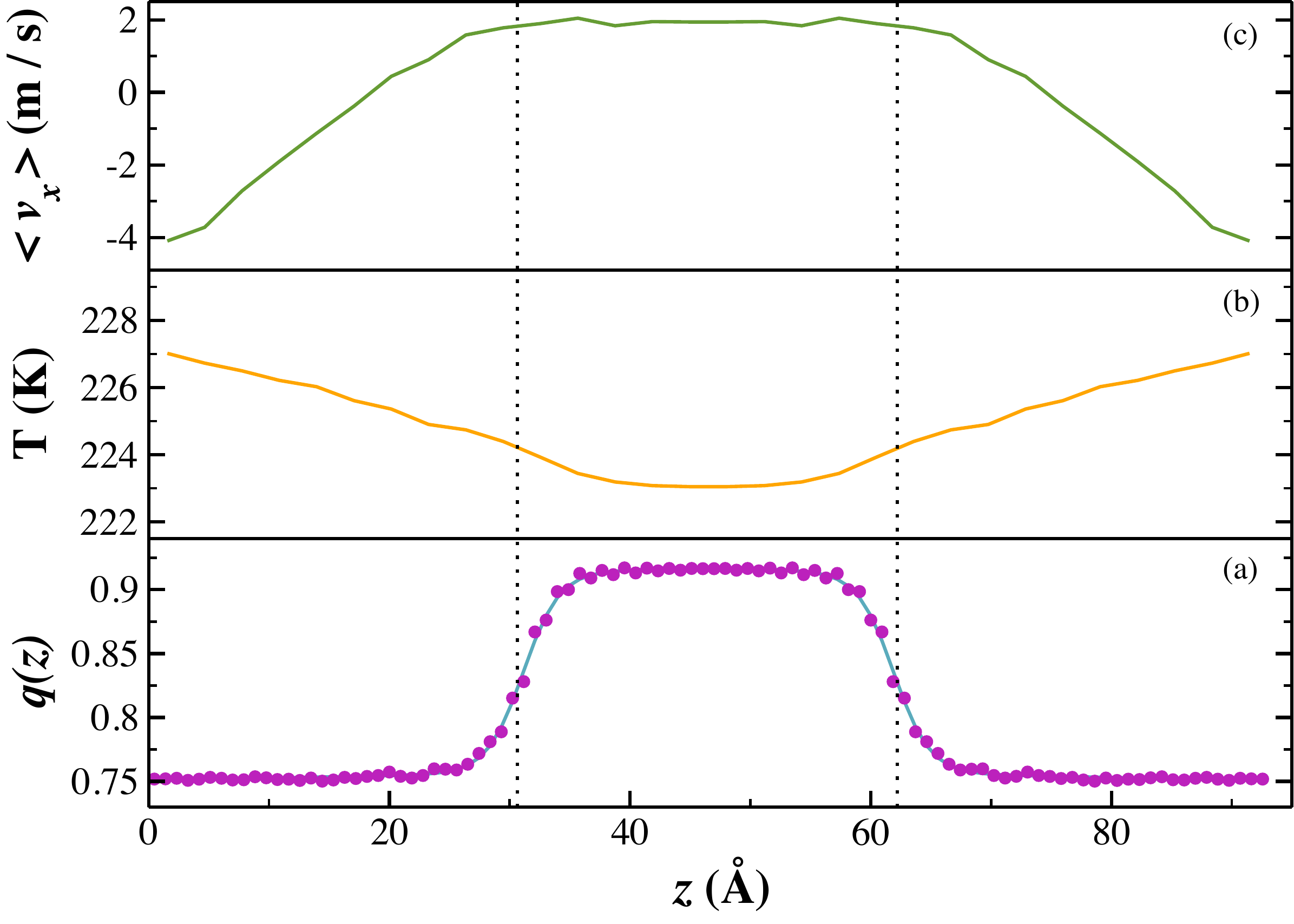}
\caption{\label{fig:pyrComic} Properties of the pyramidal interface
  being sheared through water at 3.8 ms\textsuperscript{-1}. Lower
  panel: the local tetrahedral order parameter, $q(z)$, (circles) and
  the hyperbolic tangent fit (turquoise line).  Middle panel: the
  imposed thermal gradient required to maintain a fixed interfacial
  temperature of 225 K. Upper panel: the transverse velocity gradient
  that develops in response to an imposed momentum flux. The vertical
  dotted lines indicate the locations of the midpoints of the two
  interfaces.}
\end{figure}
\newpage

\begin{figure}
\includegraphics[width=\linewidth]{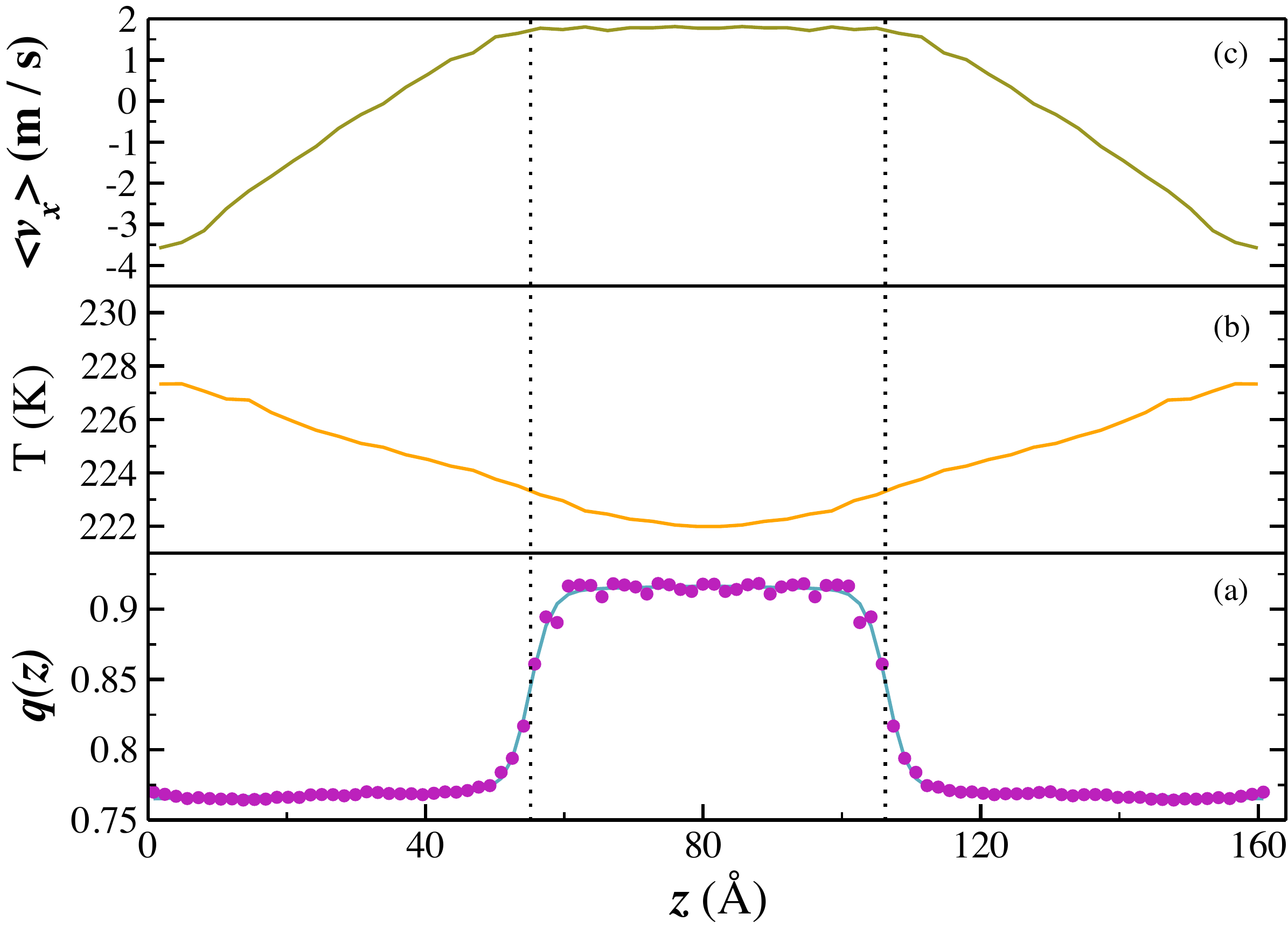}
\caption{\label{fig:spComic} The secondary prism interface with a shear 
rate of 3.5 \
ms\textsuperscript{-1}. Panel descriptions match those in figure \ref{fig:pyrComic}.}
\end{figure}
\newpage

\begin{figure}
\includegraphics[width=\linewidth]{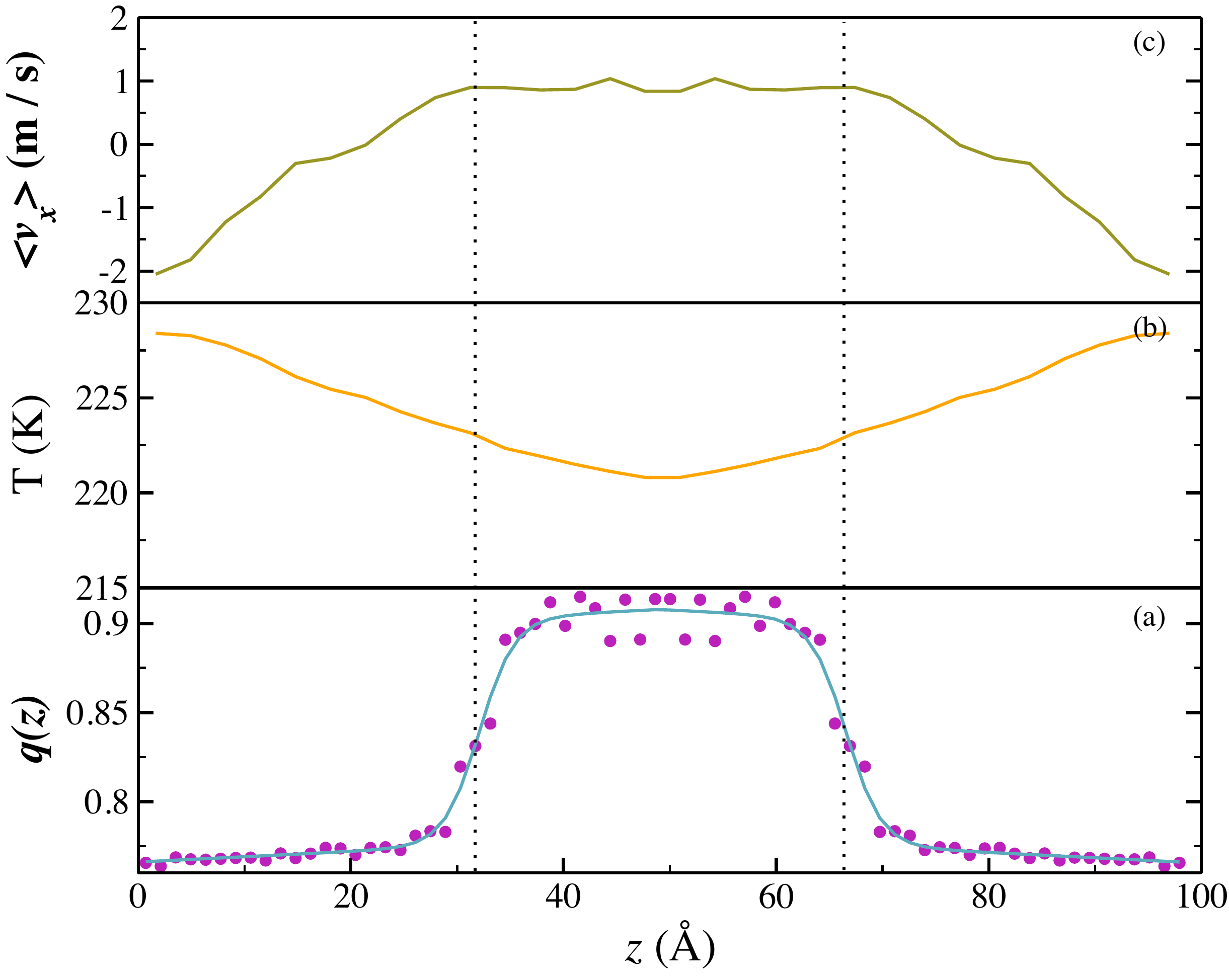}
\caption{\label{fig:bComic} The basal interface with a shear 
rate of 1.3 \
ms\textsuperscript{-1}. Panel descriptions match those in figure \ref{fig:pyrComic}.}
\end{figure}
\newpage

\begin{figure}
\includegraphics[width=\linewidth]{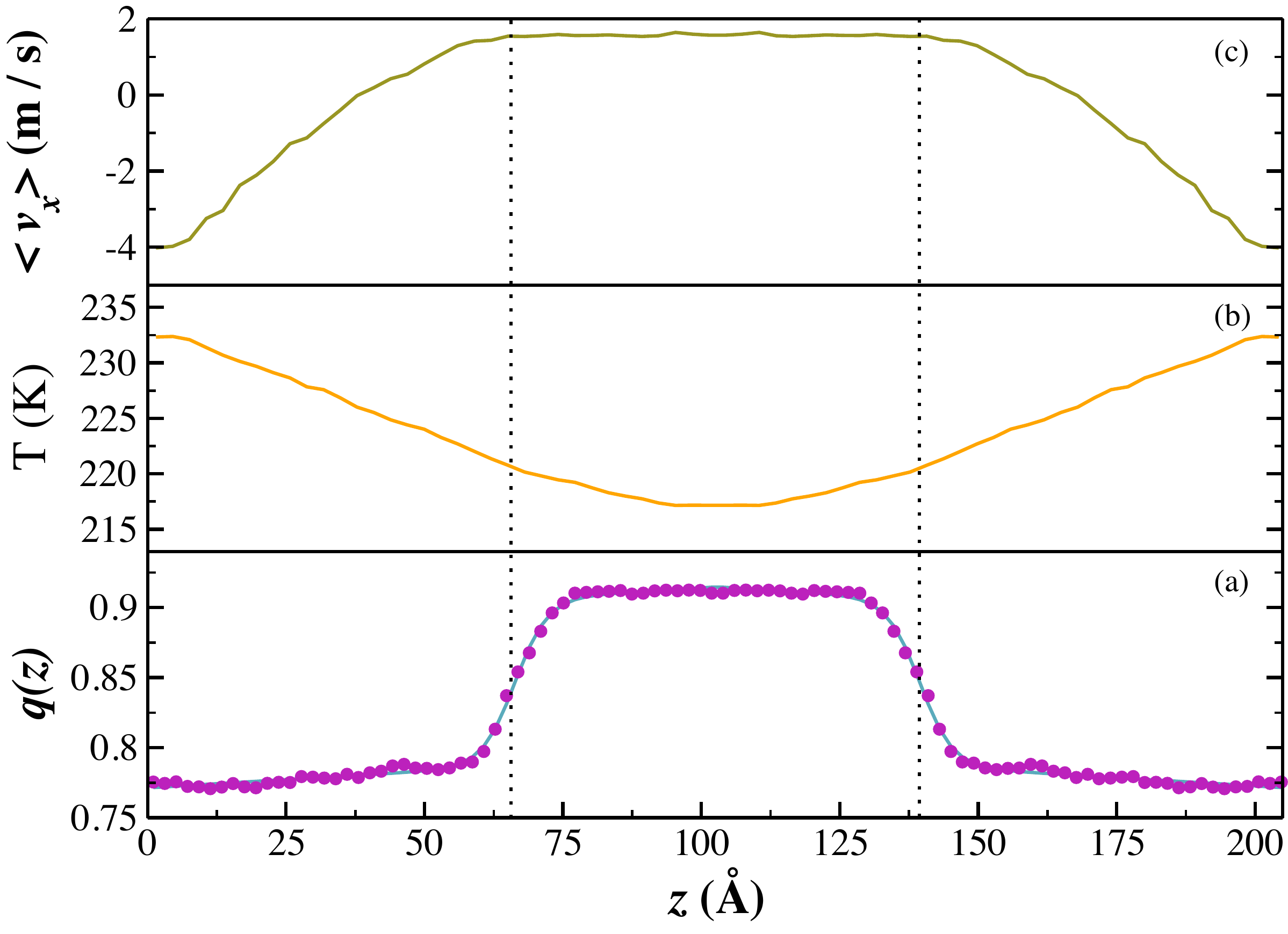}
\caption{\label{fig:pComic} The prismatic interface with a shear 
rate of 2 \
ms\textsuperscript{-1}. Panel descriptions match those in figure \ref{fig:pyrComic}.}
\end{figure}
\newpage

\begin{figure}
\includegraphics[width=\linewidth]{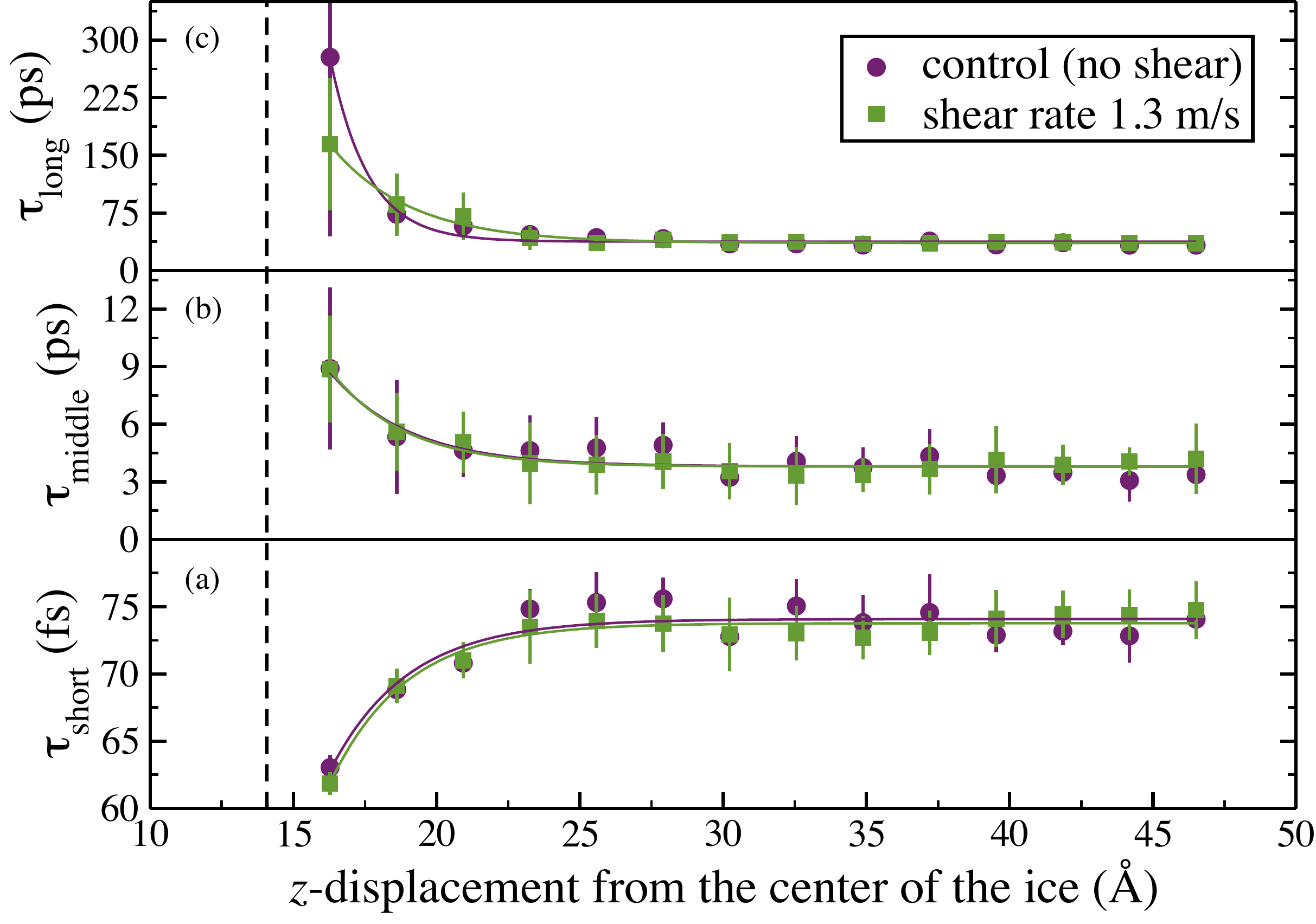}
\caption{\label{fig:PyrOrient} The three decay constants of the
  orientational time correlation function, $C_2(z,t)$, for water as a
  function of distance from the center of the ice slab. The vertical
  dashed line indicates the edge of the pyramidal ice slab determined
  by the local order tetrahedral parameter. The control (circles) and
  sheared (squares) simulations were fit using shifted-exponential
  decay (see Eq. 9 in Ref. \citealp{Louden13}).}
\end{figure}  
\newpage

\begin{figure}
\includegraphics[width=\linewidth]{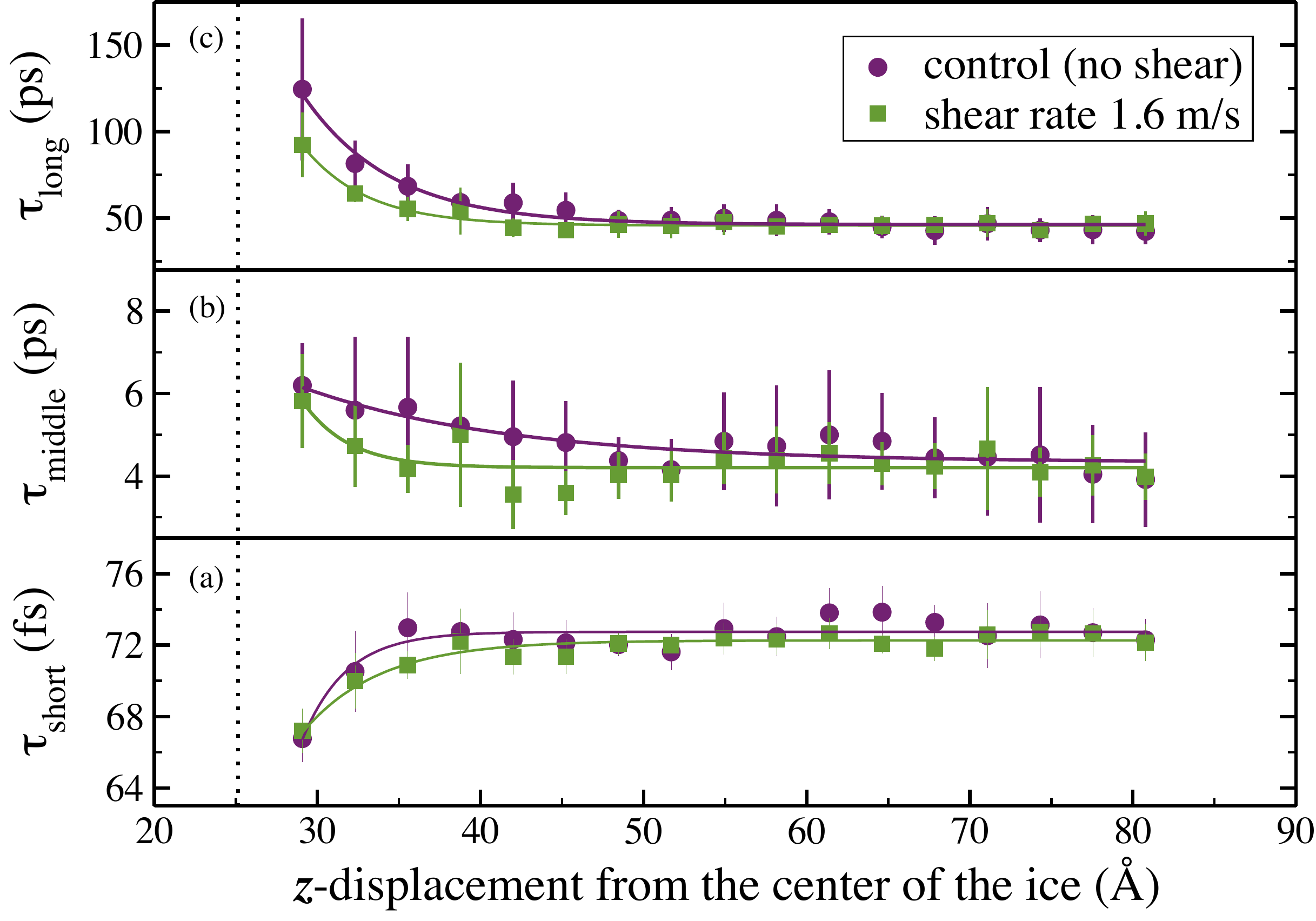}
\caption{\label{fig:SPorient} Decay constants for $C_2(z,t)$ at the secondary 
prism face. Panel descriptions match those in \ref{fig:PyrOrient}.}
\end{figure}

\newpage

\begin{figure}
\includegraphics[width=\linewidth]{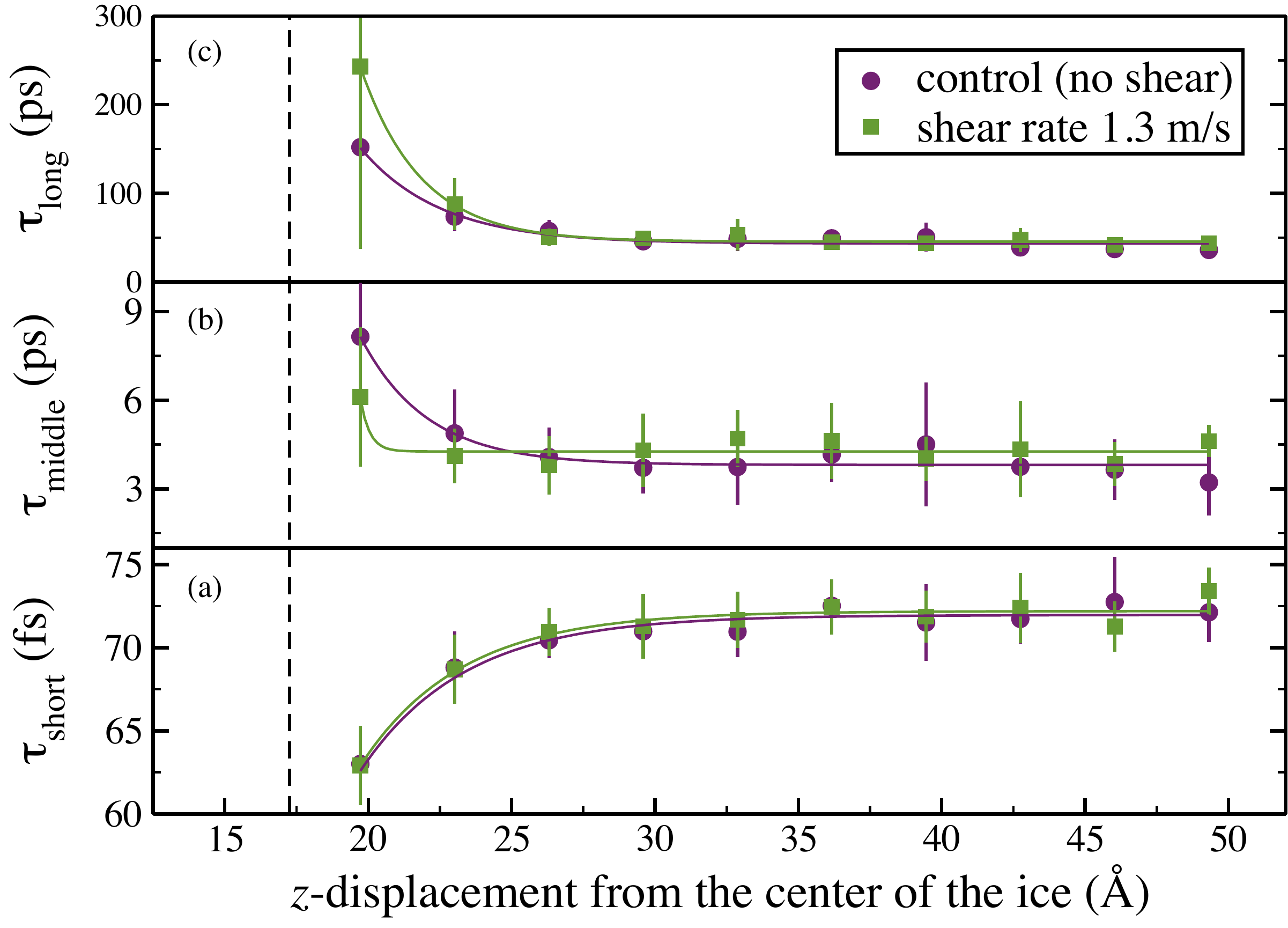}
\caption{\label{fig:Borient} Decay constants for $C_2(z,t)$ at the basal face. Panel descriptions match those in \ref{fig:PyrOrient}.}
\end{figure}
\newpage

\begin{figure}
\includegraphics[width=\linewidth]{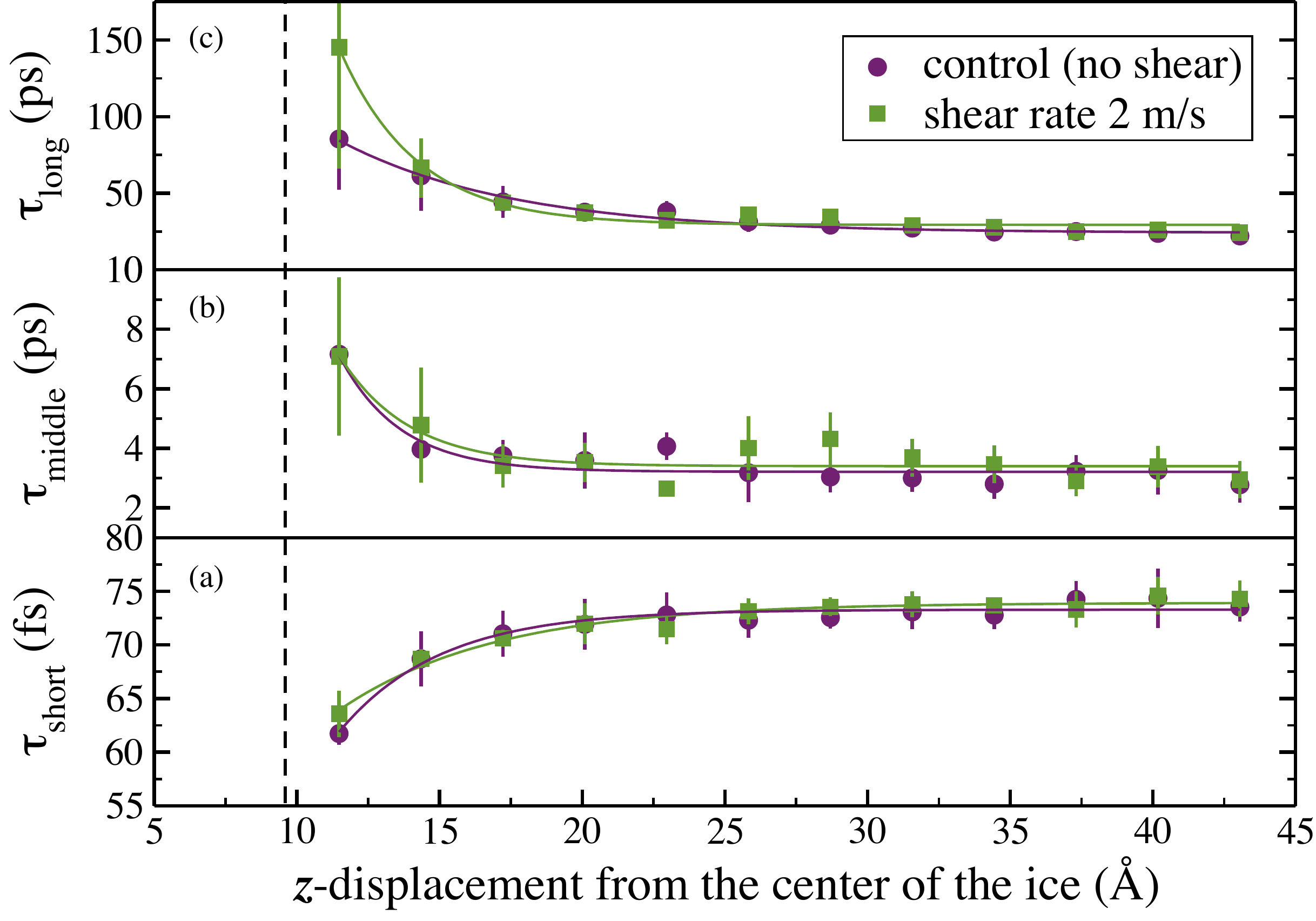}
\caption{\label{fig:Porient} Decay constants for $C_2(z,t)$ at the 
prismatic face. Panel descriptions match those in \ref{fig:PyrOrient}.}
\end{figure}